\documentclass[fullpage,12pt,superscriptaddress]{iopart}

\usepackage{graphicx}%
\usepackage{dcolumn}%
\usepackage{bm}%
\usepackage{setspace}
\usepackage{graphicx}
\usepackage{setspace}
\usepackage{bibentry}
\usepackage{dashrule}
\usepackage{color}
\usepackage{xcolor}
\usepackage{picture}
\usepackage{grffile}
  \expandafter\let\csname equation*\endcsname\relax
  \expandafter\let\csname endequation*\endcsname\relax
\usepackage{amsmath}
\usepackage{float}
\usepackage{placeins}
\usepackage{caption}
\usepackage{subcaption}
\usepackage{rotating}
\usepackage{hyperref}
\graphicspath{{../../Pictures/Two tip SECARS paper pictures_NJP/}}

\expandafter\def\expandafter\normalsize\expandafter{%
    \normalsize
    \setlength\abovedisplayskip{8pt}
    \setlength\belowdisplayskip{8pt}
    \setlength\abovedisplayshortskip{8pt}
    \setlength\belowdisplayshortskip{8pt}
}

\begin{document}

\begin{center}

\title{\Large Dual-tip-enhanced ultrafast CARS nanoscopy}


\author{\normalsize   Charles W. Ballmann\textsuperscript{1}, Bin Cao\textsuperscript{1,2}, Alexander M. Sinyukov\textsuperscript{1}, Alexei V. Sokolov\textsuperscript{1} and Dmitri V. Voronine\textsuperscript{1}\\[0.2cm]}

\address{\footnotesize \textsuperscript{1}Department of Physics and Astronomy, Texas A{\&}M University, College Station, TX 77843-4242,\\
	\textsuperscript{2}Department of Applied Physics, Xi'an Jiaotong University, Xi'an, China 710049
	}

\end{center}
\begin{abstract}
Coherent anti-Stokes Raman scattering (CARS) and, in particular, femtosecond adaptive spectroscopic techniques (FAST CARS) have been successfully used for molecular spectroscopy and microscopic imaging. Recent progress in ultrafast nano-optics provides flexibility in generation and control of optical near fields, and holds promise to extend CARS techniques to the nanoscale. In this theoretical study, we demonstrate ultrafast subwavelentgh control of coherent Raman spectra of molecules in the vicinity of a plasmonic nanostructure excited by ultrashort laser pulses. The simulated nanostructure design provides localized excitation sources for CARS by focusing incident laser pulses into subwavelength hot spots via two self-similar nanolens antennas connected by a waveguide. Hot-spot-selective dual-tip-enhanced CARS (2TECARS) nanospectra of DNA nucleobases are obtained by simulating optimized pump, Stokes and probe near fields using tips, laser polarization- and pulse-shaping. This technique may be used to explore ultrafast energy and electron transfer dynamics in real space with nanometre resolution and to develop novel approaches to DNA sequencing.  \\[0.2cm]
\end{abstract}%

\maketitle
Recent progress in nano-optics has led to many exciting applications which benefit from nanoscale subwavelength resolution such as imaging cancer cells, biosensing, and designing nanoscale devices \cite{Novotny}. Tip- and surface-enhanced techniques which use metal nanostructures offer nanoscale resolution due to their subwavelength size. It is desirable to obtain simultaneous nanometre spatial and ultrafast temporal resolution to investigate the dynamics of proteins, surface water, plasmon propagation, energy and charge transfer in biomolecules, and other ultrafast nanoscale processes \cite{Vasa}.

Raman spectroscopy provides molecular vibrational fingerprint information. In coherent Raman spectroscopy, such as coherent anti-Stokes Raman scattering (CARS)\cite{Minck,Eesley,Potma}, vibrational coherence is used to enhance molecular signals compared to weak spontaneous Raman spectroscopy. The increased sensitivity was used to develop CARS microscopy which allowed three-dimensional imaging capability\cite{Zumbusch,Kee,Silberberg}. Another approach to enhance Raman signals is using plasmonic nanoantennas for surface-enhanced Raman scattering (SERS)\cite{Fleischmann,Kneipp,Ru,Schluecker} and tip-enhanced Raman scattering (TERS)\cite{Bailo,Pettinger}. Combination of the coherence and surface enhancements may further improve sensitivity, and has been realized in surface-enhanced CARS (SECARS)\cite{Liang,Koo,Addison,Namboodiri,Wang,Steuwe,Hayazawa} and tip-enhanced CARS (TECARS)\cite{Ichimura1,Ichimura2}. Unfortunately, CARS often suffers from a large nonresonant background, but various approaches have been developed to suppress it. Optimal laser pulse shapes were designed in coherently-controlled CARS\cite{Silberberg,Dudovich,Dantus,Rhijn,Gao}, time-resolved CARS (tr-CARS)\cite{Hamaguchi,Volkmer,Pestov1,Pestov2}, and femtosecond adaptave spectroscopic techniques (FAST CARS)\cite{Scully}. Recently we demonstrated time-resolved surface-enhanced CARS (tr-SECARS) which allowed suppressing background and enhancing molecular signals using randomly aggregated gold nanoparticles \cite{Voronine}. 

In this theoretical study, we apply ultrafast near-field control to two-pulse tr-CARS using two silver tips as spatially separated optical nanoantennas. The two tips couple at different angles to a curved nanosphere waveguide and provide nanolocalized enhanced near fields for CARS (figure 1).
\begin{figure}[h]  
\centering                                                   
\includegraphics[scale=0.5,width=0.7\textwidth ,clip=true, trim= 20mm 220mm 98mm 17mm]{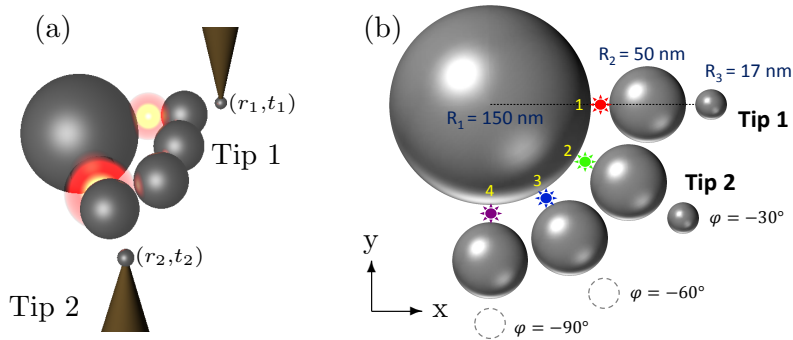}
\vspace{-5 mm}
\captionsetup{justification=raggedright,singlelinecheck=false}
\caption{\textrm{\footnotesize \setstretch{0.2} Dual-tip plasmonic nanostructure made of self-similar nanolens antennas connected by a waveguide. (a) side- and (b) top-view. The smallest nanospheres, attached to scanning probe microscope tips, provide controlled subwavelength excitation and detection. Near-field hot spots generated in the gaps between the 50 and 150 nm spheres are labelled 1--4, and highlighted by red, green, blue and purple stars, respectively. Cytosine, thymine, adenine, and guanine are placed in these hot spots, respectively. The position of tip 1 is fixed. The position of tip 2 is varied between values of the angle $\varphi $: no tip 2, $-30^\circ$, $-60^\circ$, and $-90^\circ$. The nanostructure is excited by x-polarized ultrashort laser pulses propagating along the z-direction. The dynamics of near fields is controlled by changing the position of tip 2 and by laser pulse shaping.}}
\label{setup}
\end{figure}

 This nanostructure design is analogous to plasmonic circuits made of two dipole nanoantennas coupled by a transmission line \cite{Huang1,Huang2}. Here, we model self-similar chains of silver spheres as nanoantennas \cite{Li}. The curved waveguide geometry allows compact lumping of antennas into a small region of space. Optimization of this dual-tip-enhanced CARS (2TECARS) approach requires spatiotemporal overlap of the pump, Stokes and probe fields using position-dependence of the tip enhancement and laser pulse shaping. We simulate 2TECARS spectra of DNA nucleobases placed in the gaps of the nanoantennas and demonstrate hot-spot-selective signal generation and spectral control. The dual-tip plasmonic nanostructure made of silver nanospheres is shown in figure 1. Figure 1a shows a side view with two 17 nm spheres attached to the moving tips of a scanning probe microscope. The tip mounts are made of a non-metallic material (e.g., silicon) and do not influence the nanostructure response. The structure of 50 and 150 nm spheres may be fabricated on a surface by chemical deposition with manipulation by a AFM \cite{Shafiei}. Figure 1b shows a top-view schematic (tip mounts are not shown)  with specified geometrical parameters where four 50 nm spheres are used to connect the 17 and 150 nm spheres to form self-similar nanolens antennas with orientations at $0^\circ$, $-30^\circ$, $-60^\circ$ and $-90^\circ$. These four spheres also form a curved waveguide connected to the 150 nm sphere. 

The sphere sizes were chosen according to the design of the self-similar nanolens \cite{Li}. 
The sphere radii satisfy $ R_{i+1} = \frac{R_i}{3}$. The distance between the surfaces of the consecutive spheres is $d_{i,i+1} = 0.3\,R_{i+1}$. The near-field response of the nanostructure was calculated using the multiple elastic scattering of multipole expansions (MESME) approach \cite{Garcia1,Garcia2} in the range from 200 to 700 nm with x-polarized incident plane waves for all tip 2 positions. Each incident field $E(\omega)$ was multiplied by a response function $F_{m}(\omega)$, where m corresponds to a particular tensor component. $ E_{local}=E(\omega)F_{m}(\omega)$ was used in the tip-enhanced CARS simulations. The Gaussian pump/Stokes pulses were $\sim$ 6 fs in duration corresponding to $\sim$ 2400 $\text{cm}^{-1}$ bandwidth. The Gaussian probe pulse had a bandwidth of 3 $\text{cm}^{-1}$. Pulse durations were held constant during the scan of the center frequency, with the bandwidth changing accordingly. Gaussian laser pulse shapes $E_{k}(\omega)=\mathrm {Exp}[-2\text{ln}(2)\big(\frac{\omega-\omega_{ko}}{\Delta\omega_{k}}\big)^2] ,\quad ( k=1,2 )$ were used to induce the CARS signals, where k = 1, 2 stands for pump and Stokes, respectively. 

The picosecond probe pulse was modelled by $E_{pr}(\omega) = 1,\: \text{if} \:|\omega-\omega_{pr}|\leq 1.5\,\text{cm}^{-1}\alpha,\: \text{and}\: 0$ otherwise, where $\alpha=2\pi c$, $c$ is the speed of light in vacuum, $\omega_{ko} $ is the center frequency of the  k'th pulse, and $\Delta\omega_{k}$ is the corresponding bandwidth.  The Raman data for linewidths, cross sections, and resonance frequencies was obtained for DNA nucleobases from \cite{Sigma}. The dielectric function of silver was taken from \cite{Palik}. Vacuum permittivity was used to describe the surrounding dielectric medium. Additional simulations were performed in air, but did not show significant differences. Another set of simulations was also performed using gold instead of silver. Reduced surface enhancement factors were obtained, but the CARS spectra did not show significant changes. The simulations were performed in Mathematica\textsuperscript{\textregistered} 8.

\begin{figure}
\centering
\includegraphics[scale=1,width=0.7\textwidth ,clip=true, trim= 20mm 219mm 108mm 21mm]{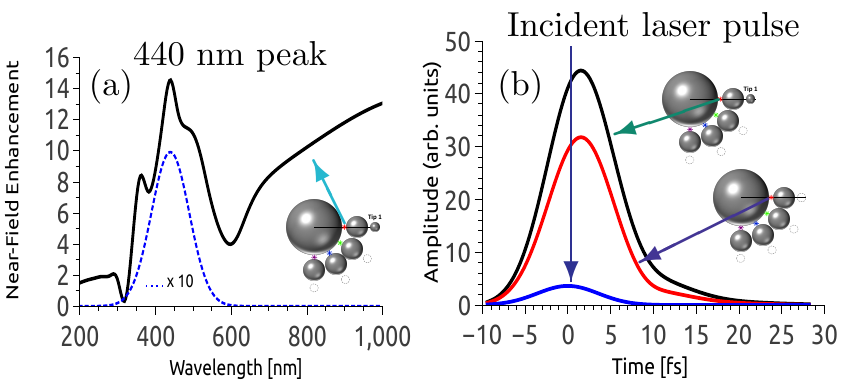}
\vspace{-0 mm}
\setstretch{0.7}\captionsetup{justification=raggedright,singlelinecheck=false}
\caption{\textrm{\footnotesize Calculated near-field enhancement in frequency and time domains. (a) x-polarized spectral response of the nanostructure without tip 2 in hot spot 1 due to excitation by the incident x-polarized ultrashort Gaussian laser pulse (dashed blue (10x)). (b) The corresponding x-polarized temporal response (black) compared to the response of a nanostructure without tips (red) and to the incident laser pulse (blue). 17 nm tip 1 provides an additional near-field enhancement.}}
\label{response at point}
\end{figure}%

\FloatBarrier

In figure 1b, four hot spots in the gaps between the 50 and 150 nm spheres are labeled 1--4 and highlighted by red, green, blue, and purple stars, respectively. Tip 1 is fixed and forms the receiving antenna for x-polarized light localized at spot 1. The x-polarized spectral response of this nanostructure without tip 2 at spot 1 is shown in figure 2a (black solid line), in which several resonances are observed with the strongest peak at 440 nm. The corresponding temporal near-field profile at spot 1 due to the excitation by x-polarized 6 fs incident laser pulses centered at 440 nm is shown in figure 2b (black line), while this response is compared to the response of a nanostructure without tips at spot 1 (red line) and to the original incident laser pulse (blue line) which shows that there is a factor of 8 enhancement of the electric field amplitude of the nanostructure without tips compared to the incident laser pulse and an additional enhancement due to the presence of tip 1. This order of magnitude field enhancement results in several orders of magnitude enhancement in the nonlinear optical signals such as CARS. The enhancement may be further improved by optimizing the nanostructure design, e.g. decreasing the gap size.

\FloatBarrier

\begin{figure} 
\centering                                                  
\includegraphics[scale=1,width=0.65\textwidth ,clip=true, trim= 20mm  175mm  108mm 14mm]{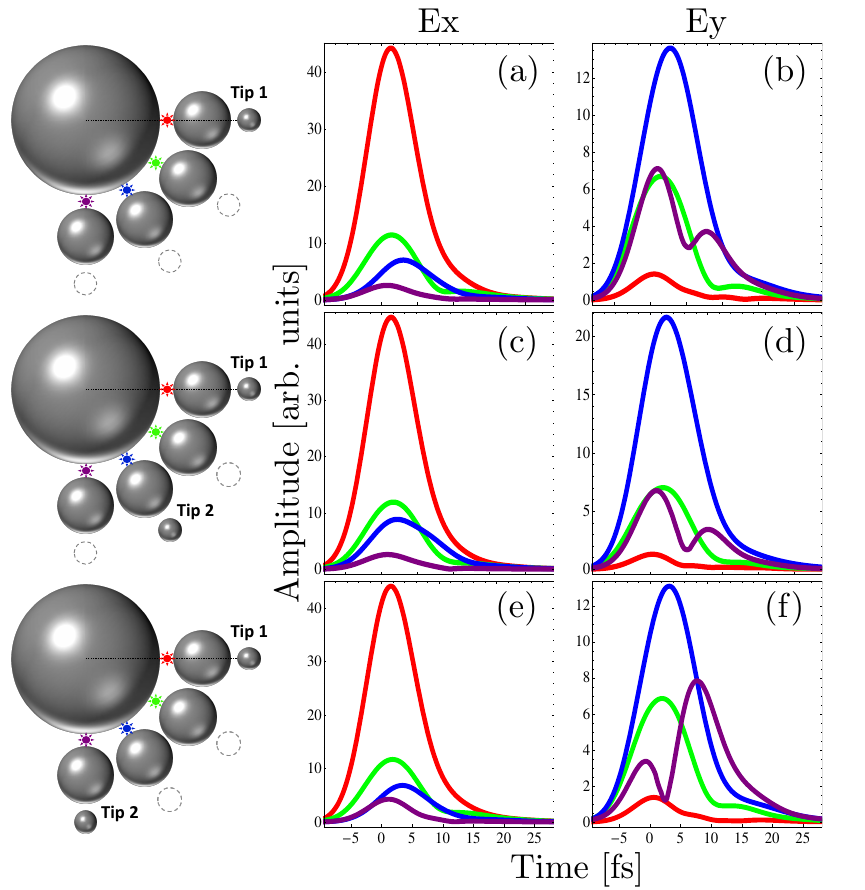}
\vspace{-5mm}
\captionsetup{justification=raggedright,singlelinecheck=false}
\caption{\textrm{\footnotesize Calculated temporal near field amplitude profiles of the pump/Stokes pulse centered at 440 nm from hot spots 1 (red), 2 (green), 3 (blue), and 4 (purple) without tip 2 (a,b) and with tip 2 at $-60^\circ$ (c,d) and $-90^\circ$ (e,f). The nanostructure geometries are shown in the left column. The x- (a,c,e) and y-polarized (b,d,f) fields are induced by x-polarized incident Gaussian 6 fs laser pulses. The pulse shapes reveal complex plasmon propagation dynamics which depend on the position of tip 2.}}
\label{strwithtime}
\end{figure}

Spatiotemporal control of the near-filed response will allow localizing and controlling CARS signals. Therefore, we next investigate the temporal near field pulse shapes from other hot spots. 
The temporal near field amplitude profiles due to the excitation by x-polarized 6 fs Gaussian laser pulses centered at 440 nm from hot spots 1--4 without tip 2 (a,b), and with tip 2 at $-60^\circ$ (c,d) and $-90^\circ$ (e,f) positions are shown in figure 3. The corresponding nanostructure geometries are sketched in the left column. The x-(Ex) and y-polarized (Ey) field profiles are shown in the middle and right columns, respectively. Complex near field temporal dynamics due to plasmon propagation and scattering is observed. Figures 3a, 3c and 3e show x-polarized fields where hot spot 1 (red) has the largest amplitude. Figures 3b, 3d and 3f show y-polarized fields where spot 3 has the largest peak amplitude. These y-polarized fields are due to depolarization of the x-polarized fields propagating along the waveguide from spot 1. Hot spot 4 becomes strongest after $\sim$ 10 fs in figure 3f. These ultrafast near field pulses are broad enough to excite vibrational coherences of typical molecules and serve as both pump and Stokes pulses in the CARS process. The temporal profile of the picosecond probe pulse is not significantly affected by the nanostructure and is not shown. However, the spectral profile of the probe is proportional to the spectral response profile such as that shown in figure 2a. The spatiotemporal overlap of the pump/Stokes and probe pulses control the intensities of CARS signals, but the near field pulses in figure 3 may also be used to trigger electronic and structural changes in molecules and to induce photochemical reactions. These pulses may be delayed in time with respect to the CARS pulses. 2TECARS may therefore be used as a nanoscopic ultrafast probing technique to monitor these processes.

The near fields are modified by the moving tips of the nanostructure and by laser pulse shaping. This provides several control parameters for manipulating CARS signals. We first use these parameters to isolate CARS nanospectra of DNA nucleobases. In our work, the third-order nonlinear polarization is modelled by
\begin{equation}
 P^{(3)}(\omega)=\int_0^\infty\chi_{R}^{(3)}(\Omega)E_{3}(\omega-\Omega)S_{12}(\Omega)\mathrm{d}\Omega,
\end{equation}
where $\chi_{R}^{(3)}$, the resonant third-order nonlinear susceptibility, is given by 
\begin{equation}
\chi_{R}^{(3)}(\omega')=\displaystyle\sum_{k} \frac{A_{k}\Gamma_{k}}{\Omega_{Rk}-\omega'-i\Gamma_{k}},
\end{equation}
and $S_{12}$ is
\begin{equation}
 S_{12}(\Omega)=\int_0^\infty E_{1}(\omega'')E_{2}^{*}(\omega''-\Omega)\mathrm{d}\omega''.
\end{equation}
$A_{k}$ is a constant related to the Raman cross-section, $\Gamma_{k} $ gives the Raman line halfwidth, and $\Omega_{Rk} $ gives the k'th vibrational frequency. In calculations with single broadband pump/Stokes laser pulses, we set $E_{1}\!=\!E_{2}\!=\!E$, with $E$ sufficiently broad to excite the desired vibrations. The non-resonant background was not included in this work. Interference between signal and background is common, but many methods have been developed to minimize the non-resonant background. Our tr-CARS approach is especially useful for background suppression\cite{Pestov1,Pestov2}.
The CARS signal is given by $I_{CARS}(\omega)\propto |P^{(3)}(\omega)|^{2}.$

\begin{figure}  
\centering                                                    
\includegraphics[scale=1,width=0.7\textwidth ,clip=true, trim= 15mm 190mm 108mm 18mm]{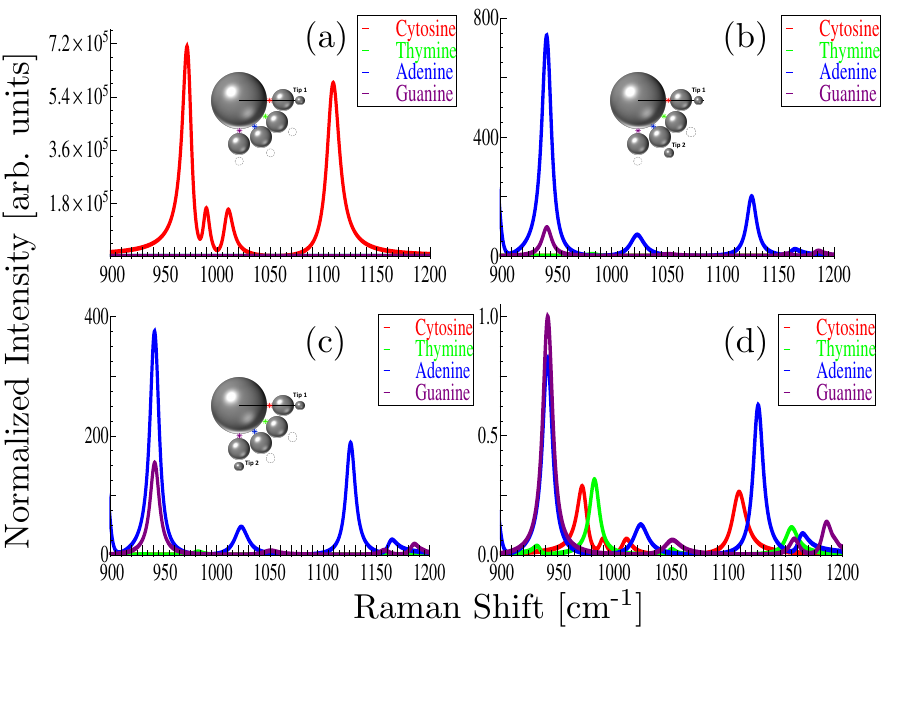}
\vspace{-10mm}
\captionsetup{justification=raggedright,singlelinecheck=false}
\caption{\textrm{\footnotesize Calculated 2TECARS nanospectra: x-polarized without tip 2 (a) and y-polarized with tip 2 at $-60^\circ$ (b) and $-90^\circ$ (c). (d) CARS spectra of nucleobases without nanostructure. Cytosine, thymine, adenine and guanine are placed in spots 1--4, and marked by red, green, blue and purple, respectively. Tip 2 is a nanoscale control "knob" for CARS spectra. The intensities are normalized to the strongest guanine peak at 942 $\text{cm}^{-1}$, and represent the enhancement factors.}}
\label{carsDNA}
\end{figure}

Figure 4d shows CARS spectra of cytosine (red), thymine (green), adenine (blue) and guanine (purple) obtained using x-polarized Gaussian 6 fs pump/Stokes (440 nm) and picosecond probe (480 nm) laser pulses without a nanostructure. All the spectra were normalized to the maximum of the strongest guanine peak at 942 $\text{cm}^{-1}$. Figures 4a--4c show CARS spectra of nucleobases placed in the hot spots of the nanostructure described above. Cytosine, thymine, adenine, and guanine were placed in spots 1, 2, 3, and 4, respectively. Figure 4a shows the x-polarized CARS spectra from the nanostructure without tip 2. CARS signal of cytosine in spot 1 dominates (red). In figure 4b, for tip 2 at $-60^\circ$, the y-polarized CARS spectra of cytosine and thymine are suppressed and spectra of adenine (blue) and guanine (purple) dominate, with adenine making the largest contribution. In figure 4c for tip 2 at $-90^\circ$, adenine (blue) dominates (y-polarized). These spectra illustrate the use of tip 2 as a nanocontrol parameter. 

To investigate the origin of the spatial dependencies of 2TECARS signals in figure 4, we performed similar simulations on pyridine molecules located at every hot spot with the pump, Stokes and probe pulses having the same parameters as in figure 4. Similar effects of tip position seen for the nucleobases were also observed for pyridine. Therefore, the effects are not simply due to different molecular responses, but are predominantly due to the structure response.

Next, we simulate spectral control of 2TECARS signals by laser polarization- and pulse-shaping. Laser pulse shaping provides additional ``knobs" for controlling ratios of spectral peaks. We achieved spectral control of CARS signals by varying the center wavelengths of the incident laser pulses in figure 5. Ratios of the y-polarized CARS signals of adenine to guanine, and of thymine to other nucleobases are maximized in figures 5a and 5b, respectively, by varying both the pump/Stokes and the probe center wavelengths. The y-polarized spectrum of guanine (purple) is amplified using the same strategy in figure 5c. For analysis, these results can be understood by examining the near field pulse shapes and overlaps via $S_{12}$ cross-correlation in (3). Figure 6 shows the pump/Stokes pulse $S_{12}(1050\: \text{cm}^{-1})$ autocorrelation plots as a function of the center wavelength for the nanostructure without tip 2 (6a), and for tip 2 positions at $-30^\circ$ (6b), $-60^\circ$ (6c) and $-90^\circ$ (6d) from different spots in the nanostructure. The average value $1050 \:\text{cm}^{-1}$ was chosen to simplify the analysis. Figure 6 reveals the optimal spectral range of the response for controlling CARS spectra by varying the pump/Stokes pulse center wavelength. For example, Figure 6c can be used to select the suitable wavelength range of the pump/Stokes pulses in order to achieve switching between adenine (figure 5a) and guanine (figure 5c) signals with tip 2 at $-60^\circ$. Figure 6c shows that the intensity (autocorrelation) of the pump/Stokes pulse is larger at spot 3 (adenine) for most of the spectral range, except in the region around 500 nm. Therefore, 332 nm and 550 nm center wavelengths were chosen to achieve the switching. Another control parameter is given by the probe pulse center wavelength and is directly proportional to the nanostructure response. Analysis of these plots allows optimizing parameters to control nanoscopic CARS signals and gives a clear understanding of the control mechanisms.

\begin{figure}   
\centering                                                     
\includegraphics[scale=1,width=0.7\textwidth ,clip=true, trim= 15mm 181mm 108mm 15mm]{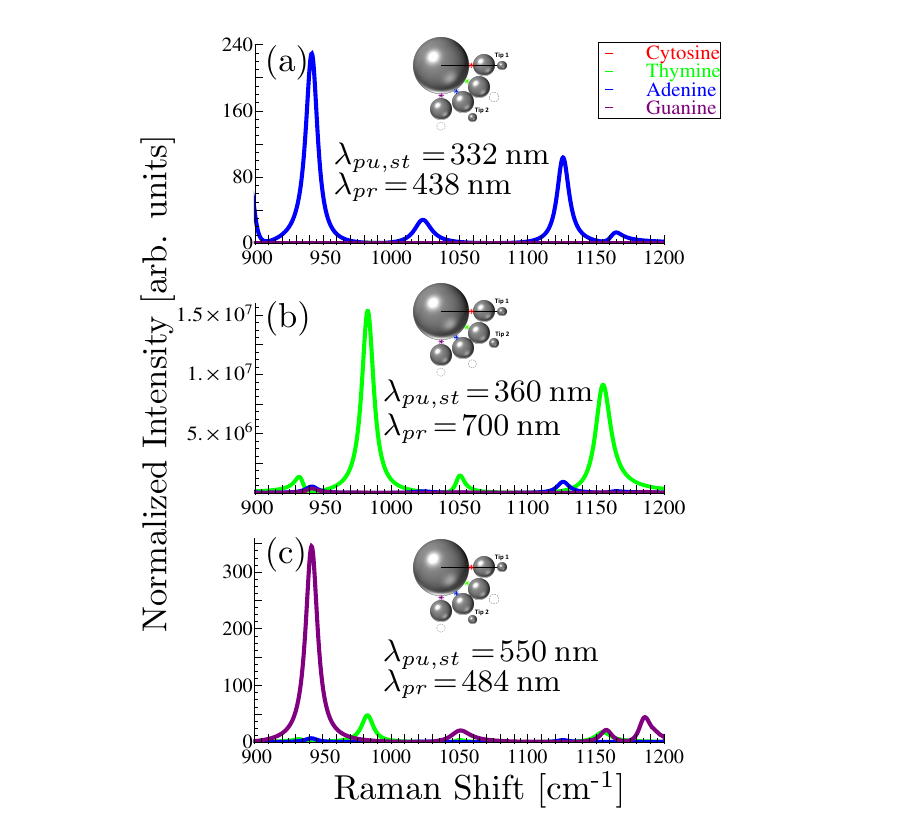}
\vspace{-5mm}
\captionsetup{justification=raggedright,singlelinecheck=false}
\caption{\textrm{\footnotesize The ratios of y-polarized spectral peaks of adenine to guanine is maximized in (a) and minimized in (c), and of thymine to other nucleobases is maximized in (b) by varying the pump/Stokes and probe pulse center wavelengths and tip 2 position. The intensities are normalized to the strongest guanine peak at 942 $\text{cm}^{-1}$ in figure 4d, and represent the enhancement factors. } }
\label{freqandtime}
\end{figure}

\FloatBarrier
\begin{figure} 
\centering                                                  
\includegraphics[scale=1,width=0.7\textwidth ,clip=true, trim= 15mm 190mm 108mm 18mm]{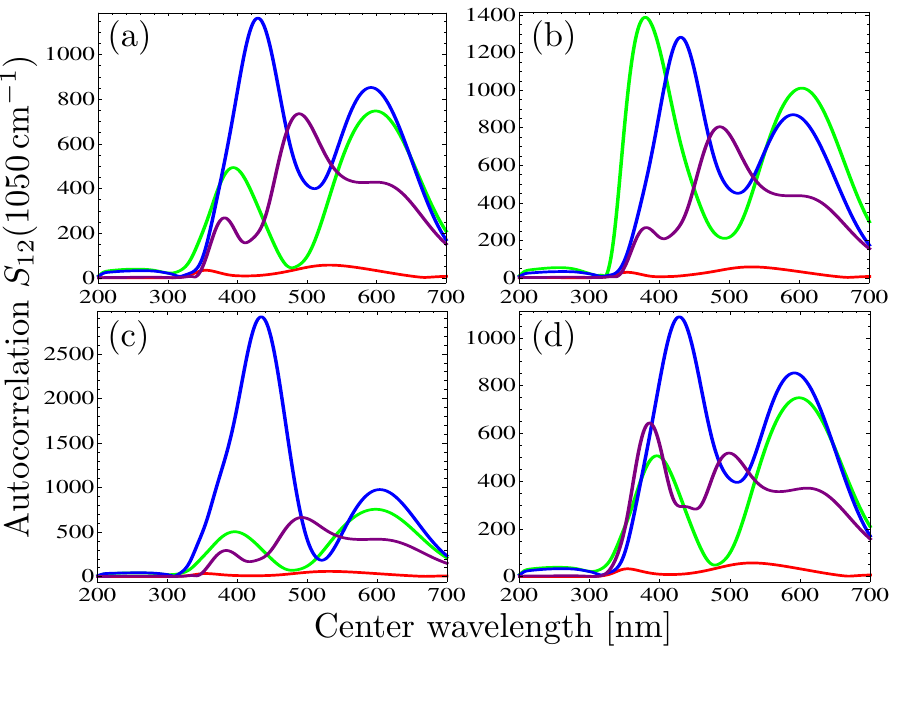}
\vspace{-10mm}
\captionsetup{justification=raggedright,singlelinecheck=false}
\caption{\textrm{\footnotesize Autocorrelation $S_{12}(1050 \:\text{cm}^{-1})$ of the y-polarized near field pump/Stokes pulses as a function of their center wavelength without tip 2 (a), and for tip 2 at $-30^\circ$ (b), $-60^\circ$ (c), and $-90^\circ$ (d) from spots 1 (red), 2 (green), 3 (blue), and 4 (purple). These plots are used to optimize ratios of 2TECARS signals from different spots and explain the results shown in figures 4--5. } }
\end{figure}

The proposed nanostructure provides a nanosphere analogue of the dual-antenna plasmonic circuit\cite{Huang1,Huang2} as an example of a controllable plasmonic system for space-time-resolved ultrafast nanoscopy. These and other geometries based on optical nanoantennas have been experimentally realized\cite{Muhlschlegel,Bharadwaj,Hulst,Biagioni,Agio}. We note that the self-similar nanolens antenna has another resonance in the gap between the 17 and 50 nm spheres which has a larger enhancement\cite{Muhlschlegel}. However, we assume that there are no molecules in that hot spot. Experimentally this may be achieved, for example, by coating only the large sphere with molecules. The nanostructure design may be further improved by using these stronger hot spots. Also, Fano resonances may provide advantages in light focusing and control\cite{Luk'yanchuk,Chang}.

Using two scanning tips and pulse shaping, the CARS signal enhancement factors (EFs) of up to $10^7$ were obtained (figure 5b). This corresponds to the expectations based on the $\sim 10^1$ near field enhancement in the hot spots using (1)--(3). In this letter, we focus on separating CARS signals from different hot spots and on increasing the contrast rather that optimizing EFs. In principle, the EF for TECARS can reach $\sim 10^{18}$ for the strongest hot spots with the near fields enhanced by $\sim 10^3$ in the self-similar nanolens antenna\cite{Li}. CARS signals may be further increased by coupling two nanolens antennas. For example, we obtain an order of magnitude increase in EF by adding the second tip (figure 5b compared to figure 4a). In addition, the results presented here can be generalized to remove one or both tips by building the nanostructures with the small spheres on the substrate, with all different arrangements manufactured on one surface.

Coherent control\cite{Rice,Shapiro} by phase- and amplitude- laser pulse shaping\cite{Weiner} may be used to control spatiotemporal plasmon dynamics in nanostructures\cite{Stockman,Aeschlimann,Aeschlimann3,Aeschlimann4,Tuchscherer} and may further improve the EFs and performance of ultrafast nanoscopic space-time-resolved spectroscopy\cite{Brixner}. This approach may be applied to other nonlinear optical techniques\cite{Mukamel} such as surface-enhanced four-wave mixing\cite{Genevet}, and coherent two-dimensional nanoscopy\cite{Aeschlimann2}. Thus, multiparameter optimization may improve EFs and contrast of nonlinear signals, and will be considered in future work.

In conclusion, we propose a new approach to probing ultrafast nanoscale phenomena using ultrafast 2TECARS nanoscopy. Dual-tip-enhanced coherent Raman spectra of DNA nucleobases separated on a nanometre scale are obtained using a combination of two scanning tips and laser pulse shaping. This technique provides useful control knobs for manipulating CARS nanospectra and will further advance the field of nanobiophotonics.

We acknowledge the support of the Office of Naval Research and the Welch Foundation Grant No. A-1547. C.W.B. is supported by the Herman F. Heep and Minnie Belle Heep Texas A$\&$M University Endowed Fund held/administered by the Texas A$\&$M Foundation.

\end{document}